\documentclass[pre, aps, showpacs, showkeys]{revtex4}
\usepackage{graphicx}
\begin{document}
%\draft

\title{Theory of solitary waves in complex plasma
lattices \footnote{Proceedings of the \textit{International
Conference on Plasma Physics - ICPP 2004}, Nice (France), 25 - 29
Oct. 2004; contribution P3-042; available online at:
\texttt{http://hal.ccsd.cnrs.fr/ccsd-00001890/en/}  .}}
% Force line breaks with \\

\author{Ioannis Kourakis\footnote{On leave from: U.L.B. -
Universit\'e Libre de Bruxelles, Physique Statistique et Plasmas
C. P. 231, Boulevard du Triomphe, B-1050 Brussels, Belgium; also:
Facult\'e des Sciences Apliqu\'ees - C.P. 165/81 Physique
G\'en\'erale, Avenue F. D. Roosevelt 49, B-1050 Brussels, Belgium;
\\Electronic address: \texttt{ioannis@tp4.rub.de}},
Padma Kant Shukla\footnote{Electronic address:
\texttt{ps@tp4.rub.de}} and Bengt Eliasson\footnote{Electronic
address: \texttt{bengt@tp4.rub.de}}} \affiliation{Institut f\"ur
Theoretische Physik IV, Fakult\"at f\"ur Physik und Astronomie,
Ruhr--Universit\"at Bochum, D-44780 Bochum, Germany}
\date{\today}

\begin{abstract}
A comprehensive analytical theory for nonlinear excitations
related to horizontal (longitudinal, acoustic mode) as well as
vertical (transverse, optical mode) motion of charged dust grains
in a dust crystal is presented. Different types of localized
excitations, similar to those well known in solid state physics,
are reviewed and conditions for their occurrence and
characteristics in dusty plasma crystals are discussed. By
employing a continuum approximation  (i.e. assuming a long
variation scale, with respect to the inter-particle distance) a
dust crystal is shown to support nonlinear kink-shaped supersonic
solitary excitations, associated with longitudinal dust grain
displacement, as well as modulated envelope localized modes
associated with either longitudinal or transverse oscillations.
Although a one-dimensional crystal is considered for simplicity,
the results in principle apply to a two-dimensional lattice if
certain conditions are satisfied. The effect of mode-coupling is
also briefly considered. The relation to previous results on
atomic chains, and also to experimental results on
strongly-coupled dust layers in gas discharge plasmas, is briefly
discussed.
\end{abstract}


\maketitle

\section{Introduction}

Dust contaminated plasmas (dusty plasmas, DP) have been attracting
significant interest recently. Particularly important are dust
quasi-lattices, which are typically formed in the sheath region
above the negative electrode in discharge experiments,
horizontally suspended at a levitated equilibrium position at $z =
z_0$, where gravity and electric (and/or magnetic) forces balance.
The linear regime of low-frequency
%dust grain
oscillations in DP crystals, in the longitudinal (acoustic mode)
and transverse (in-plane, shear acoustic mode and vertical,
off-plane optical mode) direction(s), is now quite well
understood. However, the \emph{nonlinear} behaviour of DP crystals
is still mostly unexplored, and has lately attracted experimental
[1 - 3] and theoretical  [1 - 9] interest.

Recently [5], we considered the coupling between the horizontal
($\sim \hat x$) and vertical (off-plane, $\sim \hat z$) degrees of
freedom in a dust mono-layer;
%for (arbitrary) inter-grain interaction $U(r)$
%(e.g. Debye or else)
%and sheath $\Phi(z)$
%(not necessary parabolic)
%potential form(s);
a set of
%(coupled)
nonlinear equations for longitudinal and transverse dust lattice
waves (LDLWs, TDLWs) was thus rigorously derived [5]. Here, we
review the nonlinear dust grain excitations which may occur in a
DP crystal (here assumed quasi-one-dimensional and infinite,
composed from identical grains, of equilibrium charge $q$ and mass
$M$, located at $x_n = n\, r_0 , \,$\ $n \in \cal N$).
%, by explicitly assuming Debye-type interactions $U_D(r)$ and a
%(non-harmonic)
%on-site potential in the $z-$direction, viz. $\Phi(z)$.
Ion-wake and ion-neutral interactions (collisions) are omitted, at
a first step.
%, in this simplified model.
This study complements recent experimental investigations [1-3]
and may hopefully motivate future ones.

\section{\emph{Transverse} envelope structures.} 
\begin{figure}[htb]
 \centering
 \resizebox{12cm}{!}{
 \includegraphics[]{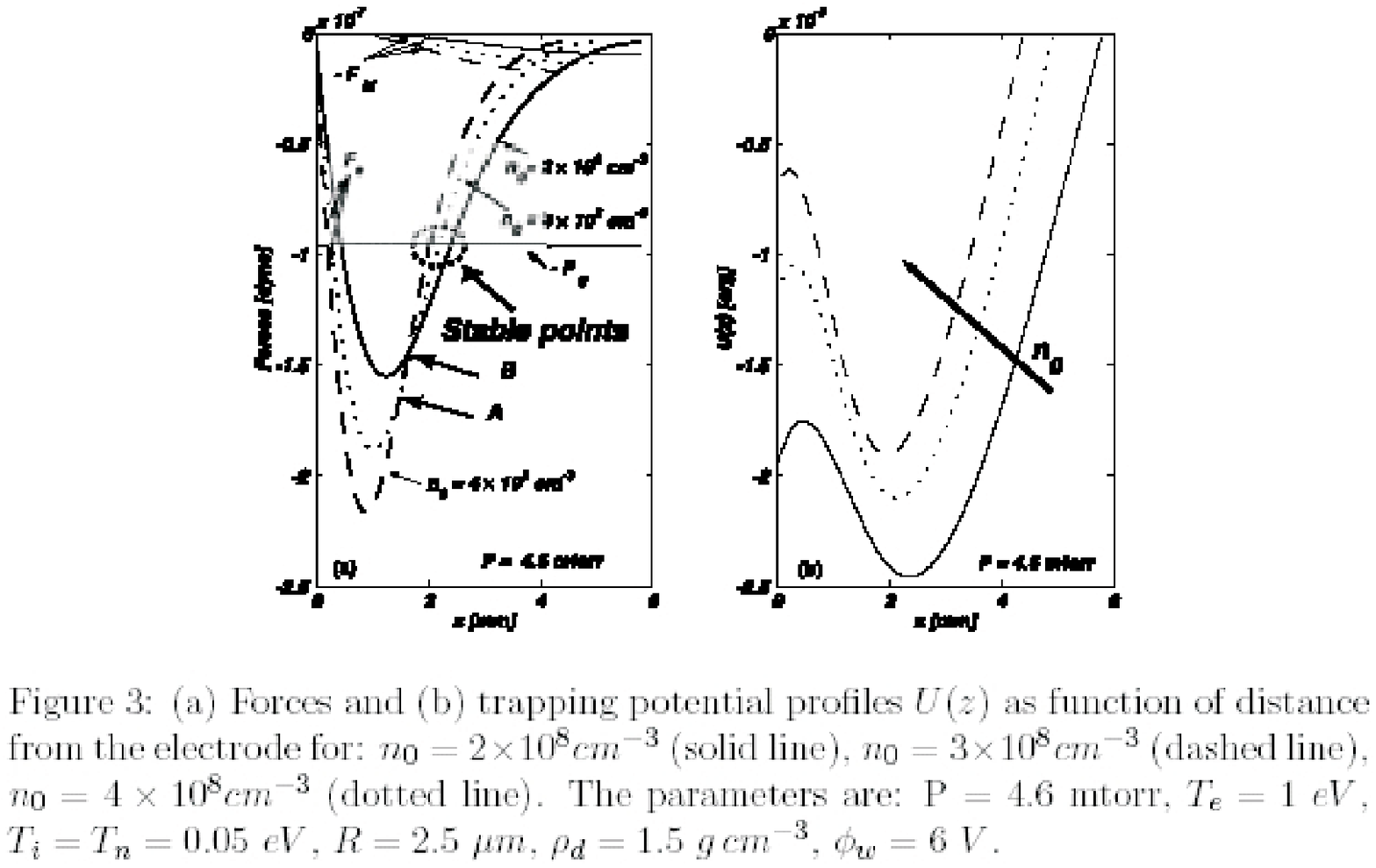}
} \caption{\small{The (anharmonic) sheath (a) force $F(z)$, and (b) force potential $V(z)$, depicted vs. the vertical distance $z$ from the negative electrode, in plasma discarge experiments; figure reprinted from
\cite{Sorasio}. }} \label{GFfig1}
\end{figure}

The vertical
(off-plane) $n-$th grain displacement $\delta z_n = z_n - z_0$ in
a dust crystal
%where $n= ..., -1, 0, 1, 2, ...$),  taking into account the intrinsic
%nonlinearity of the sheath electric (and/or magnetic) potential.
%The vertical grain displacement
obeys the equation \cite{comment1, comment2}
%of the form
\begin{eqnarray}
\frac{d^2 \delta z_n}{dt^2} + \nu \, \frac{d (\delta z_n)}{dt} +
\, \omega_{T, 0}^2 \, (\,\delta z_{n+1} + \,\delta z_{n-1} - 2
\,\delta z_n)
%\qquad \nonumber \\ \qquad
+ \omega_g^2 \, \delta z_n + \alpha \, (\delta z_n)^2 + \beta \,
(\delta z_n)^3  = 0 \, . \label{eqmotion}
\end{eqnarray}
%where $\delta z_n = z_n - z_0$ denotes the small displacement of
%the $n-$th grain around the (levitated) equilibrium position
%$z_0$, in the transverse ($z-$) direction.
The characteristic frequency \[\omega_{T, 0} \,  = \bigl[ - q
U'(r_0)/(M r_0) \bigr]^{1/2}\]
is related to the
%(electrostatic)
interaction potential  $U(r)$ [e.g. for a Debye-H\"uckel
potential: \( U_D(r) = ({q}/{r}) \,e^{-{r/\lambda_D}} \, ,\) one
has
\begin{equation} \omega_{0,
D}^2\,  = \omega_{DL}^2 \,\exp(-\kappa)\, (1 + \kappa)/\kappa^3 \,
, \label{Debye-frequency}
\end{equation} where $\omega_{DL} = [q^2/(M \lambda_D^3)]^{1/2}$ is the
characteristic dust-lattice frequency scale; $\lambda_D$ is the
Debye length; $\kappa = r_0/\lambda_D$ is the DP lattice
parameter].
%\cite{psbook}.
The \emph{gap frequency} $\omega_g$ and the nonlinearity
coefficients $\alpha, \beta$ are defined via the potential
\begin{equation}\Phi(z) \approx \Phi(z_0) + M \biggl[\frac{1}{2} \omega_g^2 \delta z_n^2
 + \frac{\alpha}{3}
\, (\delta z_n)^3 + \frac{\beta}{4}  \, (\delta z_n)^4 \biggr] \,
+ {\cal O}[(\delta z_n)^5]\end{equation} (formally expanded near
$z_0$, taking into account the electric and/or magnetic field
inhomogeneity and charge variations \cite{comment3}), i.e. leading
to an overall vertical force
\[F(z) = F_{el/m}(z) - Mg \equiv - \partial \Phi(z)/\partial z
\approx - M [\omega_g^2 \delta z_n + \alpha \, (\delta z_n)^2 +
\beta \, (\delta z_n)^3 ] \, + {\cal O}[(\delta z_n)^4]\, .\]
Recall that $F_{e/m}(z_0) = M g$. Notice the difference in
structure from the usual nonlinear Klein-Gordon equation used to
describe 1d one-dimensional oscillator chains:
%: -- cf. e.g. Eq. (1) in Ref. \cite{Kivshar}:
TDLWs (\emph{`phonons'}) in this chain are stable \emph{only} in
the presence of thanks to the field force $F_{e/m}$ (via
$\omega_g$). It should be stressed that the validity of this anharmonicity hypothesis is indeed suggested real discharge experiments, in particular for low pressure and/or density values, and also confirmed by ab initio models \cite{Sorasio} (see Fig. \ref{GFfig1}).

Linear transverse dust-lattice excitations,
%(phonons),
viz. $\delta z_n \sim \cos \phi_n$ (here $\phi_n = n k r_0 -
\omega t$) obey the \emph{optical-}like \emph{discrete} dispersion
relation \cite{comment4}:
\begin{equation}
\omega^2\,  = \omega_g^2\, - 4 \omega_{T, 0}^2\, \sin^2 \bigl( {k
r_0}/{2} \bigr)  \equiv \omega_T^2\, . \label{dispersion-discrete}
\end{equation}
%The wave frequency $\omega$ decreases with increasing wavenumber
%$k = 2 \pi/\lambda$ (or decreasing wavelength $\lambda$), implying
%that t
\begin{figure}[htb!]
 \centering
 \resizebox{6.5cm}{!}{
 \includegraphics[]{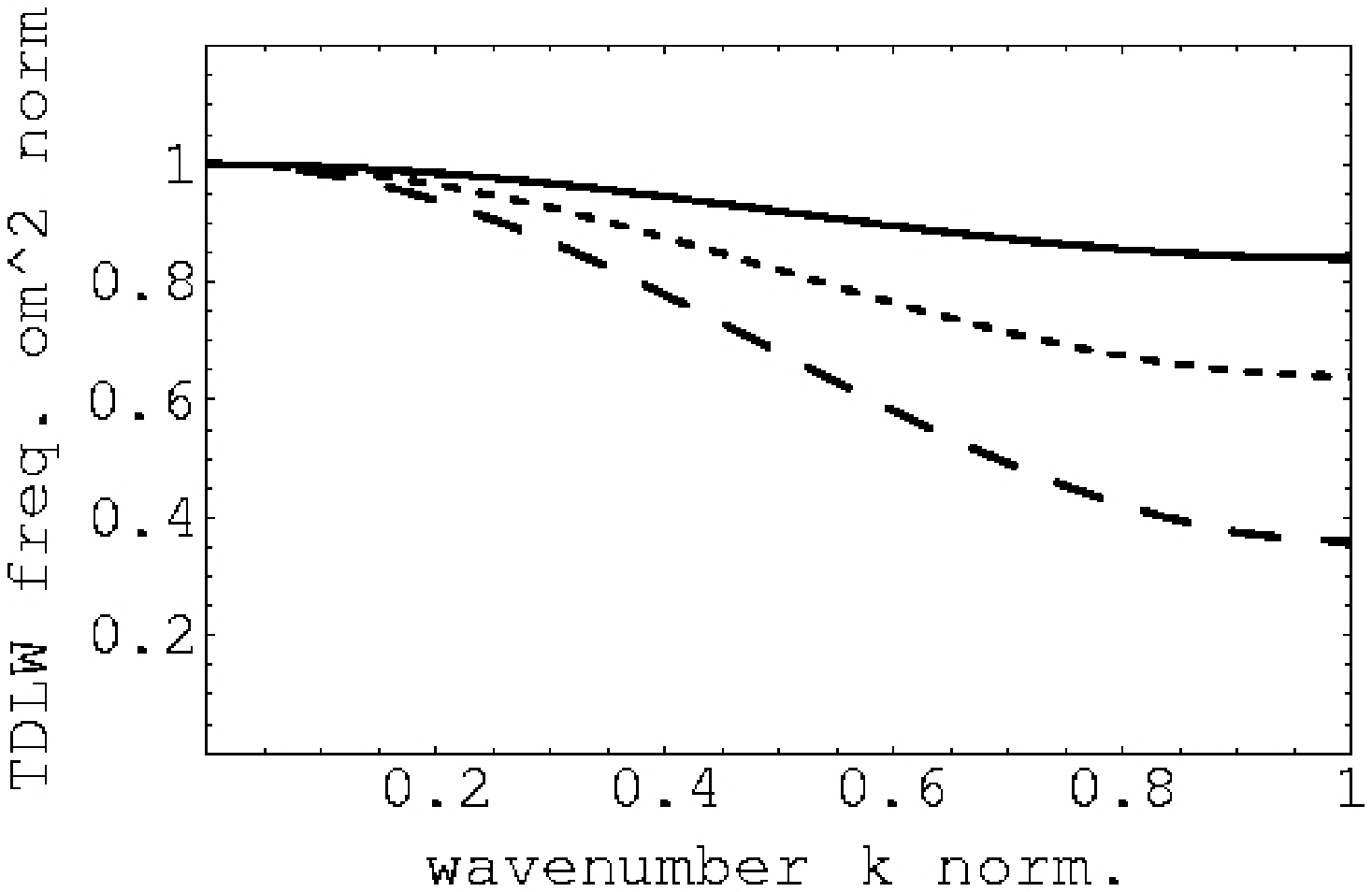}
} \caption{\small{The TDLW dispersion relation: frequency (square) $\omega_T^2$ vs. wavenumber $k$.}} \label{dispTDLW}
\end{figure}
The TDLW dispersion curve is depicted in Fig \ref{dispTDLW}.
Transverse vibrations propagate as a \emph{backward wave} [see
that $v_{g, T} = \omega_T'(k) < 0$] --
%in fact regardless of the
for any form of $U(r)$ --
%\footnote{\ See that this fact does not depend on one's choise of
%interaction potential $U(r)$.},
%: the group velocity $v_g = \omega'(k)$ and the phase speed
%$v_{ph}=\omega/k$ have opposite directions (this is
%in agreement with
cf. recent experiments [2].
%In Eq. (\ref{dispersion-discrete}) w
Notice
%the \emph{gap frequency} $\omega_g$, as well as
the lower cutoff $\omega_{T, min}\, = (\omega_g^2\, - 4 \omega_{T,
0}^2)^{1/2}$ (at the edge of the Brillouin zone, at $k =
\pi/r_0$), which is \emph{absent in the continuum limit}.
%, viz. $\omega^2\, \approx \omega_g^2\, - \omega_0^2 \, k^2 \, r_0^2$
(for $k \ll r_0^{-1}$).

Allowing for a slight departure from the small amplitude (linear)
assumption, one obtains:
%\(\delta z_n \approx \epsilon \,
%(w_1^{(1)} \, e^{i \phi_n} + {\rm{c.c.}} ) \, + \,  \epsilon^2 \,
%\bigl[w_0^{(2)} + \, (w_2^{(2)} \, e^{2 i \phi_n} +
%{\rm{c.c.}})\bigr]\, + ... \, \)
\begin{equation}
\delta z_n \approx \epsilon \, (A \, e^{i \phi_n} + {\rm{c.c.}}) +
\epsilon^2 \, \biggl[-\frac{2 |A|^2}{\omega_g^2} + \,
\biggl(\frac{A^2}{3 \omega_g^2} \, e^{2 i \phi_n} + {\rm{c.c.}}
\biggr)\biggr] \, + ... \, . \end{equation}
%(where $w_0^{(2)}\sim |A|^2$, $w_2^{(2)} \sim A^2$);
Notice the generation of higher phase harmonics due to
nonlinearity. The (slowly varying) amplitude $w_1^{(1)} \equiv
A[\epsilon (x - v_g t), \epsilon^2 t]$ obeys a {\em nonlinear
Schr\"{o}dinger equation} (NLSE) in the form [7]:
\begin{equation}
i\, \frac{\partial A}{\partial T} + P\, \frac{\partial^2
A}{\partial X^2} + Q \, |A|^2\,A = 0 \, , \label{NLSE}
\end{equation}
where $\{ X, T \}$ are the \emph{slow} variables $\{ \epsilon (x -
v_g t), \epsilon^2 t \}$. The {\em dispersion coefficient} $P$ is
related to the curvature of $\omega(k)$ as
 $P_T = \omega_T''(k)/2$ is negative/positive for low/high
values of $k$. The {\em nonlinearity coefficient} \begin{equation}
Q = \frac{1}{2 \omega_T} \biggl( \frac{10 \alpha^2}{3 \omega_g^2}
- 3\, \beta \biggl)
\end{equation} is positive for \emph{all} known experimental
values of the anharmonicity coefficients $\alpha$, $\beta$ [3].
%Without going into many details, let us
For long wavelengths [i.e. $k < k_{cr}$, where $P(k_{cr})=0$], the
theory [7] predicts that TDLWs will be  modulationally stable, and
may propagate in the form of dark/grey envelope excitations
(\emph{hole} solitons or \emph{voids}; see Fig. \ref{fig1}a,b). On
the other hand, for $k > k_{cr}$, \emph{modulational instability}
may lead to the formation of bright (\emph{pulse}) envelope
solitons (see Fig. \ref{fig1}c). Analytical expressions for these
excitations can be found in [7].
\begin{figure}[htb]
 \centering
 \resizebox{13cm}{!}{
 \includegraphics[]{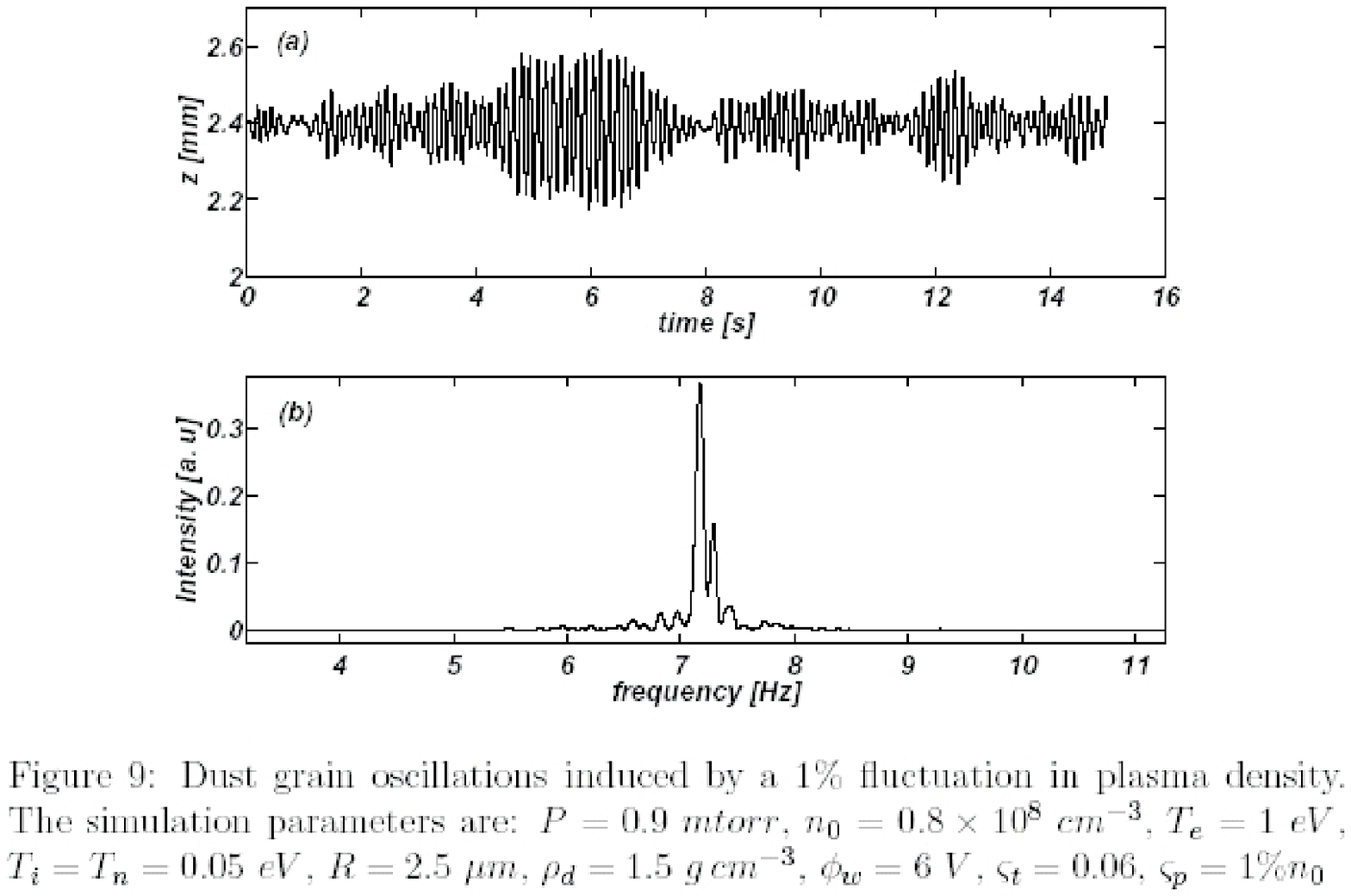}
} \caption{\small{Amplitude modulation of transverse dust lattice oscillations; simulation data provided in the embedded caption; figure reprinted from
\cite{Sorasio}. }} \label{GFfig2}
\end{figure}
\begin{figure}[htb]
 \centering
 \resizebox{14.6cm}{!}{
 \includegraphics[]{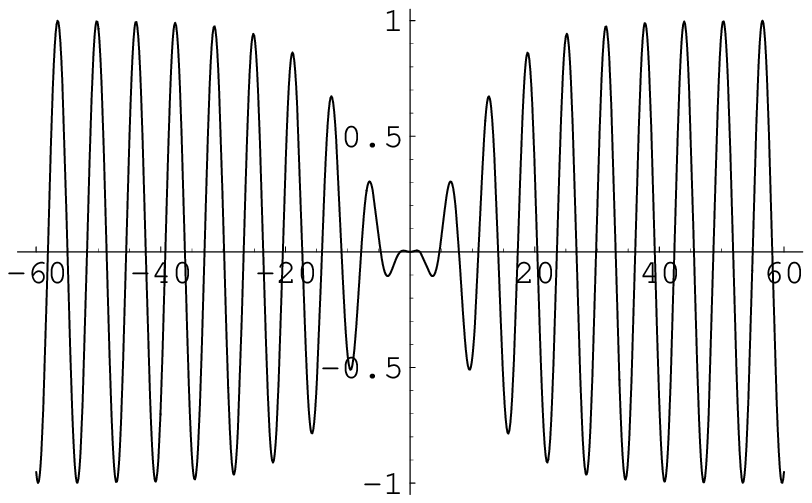}
\hskip .5 cm
\includegraphics{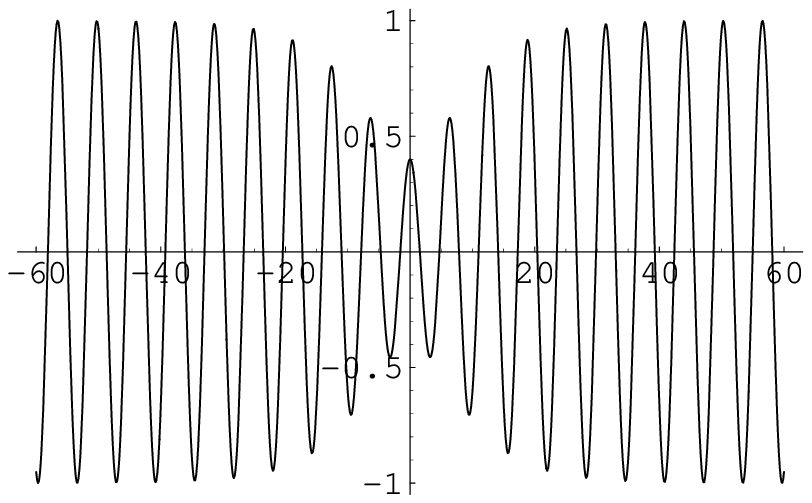}
\hskip .5 cm
\includegraphics{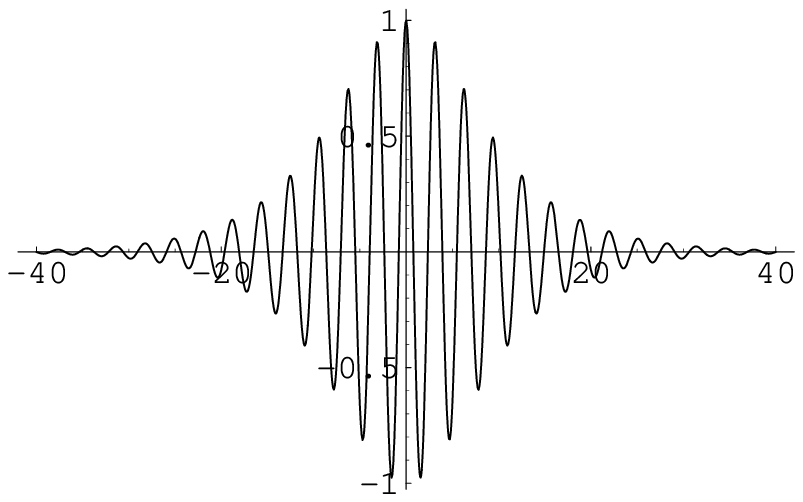}
} \caption{\small{TDL
%packets
%modulated wavepackets
envelope solitons of the (a) \emph{dark}, (b) \emph{grey}, and (c)
\emph{bright} type.}} \label{fig1}
\end{figure}

It may be noted that the modulation of transverse dust grain oscillations clearly appears in numerical simulations \cite{Sorasio}; see e.g Fig.
\ref{GFfig2}.

\newpage

\section{\emph{Longitudinal} envelope  excitations.}

The longitudinal dust grain displacements $\delta x_n = x_n - n
r_0$ are described by the \emph{nonlinear} equation of motion
\cite{IKPKSLDLW, comment1}:
\begin{eqnarray}
\frac{d^2 (\delta x_n)}{dt^2} \, + \nu \, \frac{d (\delta
x_n)}{dt}  = \, \omega_{0, L}^2 \, (\delta x_{n+1} + \delta
x_{n-1} - 2 \delta x_{n}) \qquad  \qquad   \qquad  \qquad   \qquad
\qquad \qquad \qquad
\nonumber \\
 \, - a_{20}
\, \bigl[ (\delta x_{n+1} - \delta x_{n})^{2} -  (\delta x_{n} -
\delta x_{n-1})^{2} \bigr]
%\nonumber \\
 \, + \, a_{30}  \, \bigl[ (\delta
x_{n+1} - \delta x_{n})^{3} -  (\delta x_{n} - \delta x_{n-1})^{3}
\bigr] \, .
 \label{discrete-eqmotion-x}
\end{eqnarray}
The resulting linear mode \cite{comment4} obeys the
\emph{acoustic} dispersion relation: \begin{equation} \omega^2\, =
4 \omega_{L, 0}^2\, \sin^2 \bigl( {k r_0}/{2} \bigr) \equiv
\omega_L^2 \, ,
\end{equation}
 where $\omega_{L, 0} =
[U''(r_0)/M)]^{1/2}$; in the Debye case, ${\omega_{L, 0}^2} = 2 \,
\omega_{DL}^2 \, \exp({-\kappa}) \, (1 + \kappa +
\kappa^2/2)/{\kappa^3} $.
\begin{figure}[htb]
 \centering
 \resizebox{7cm}{!}{
 \includegraphics[]{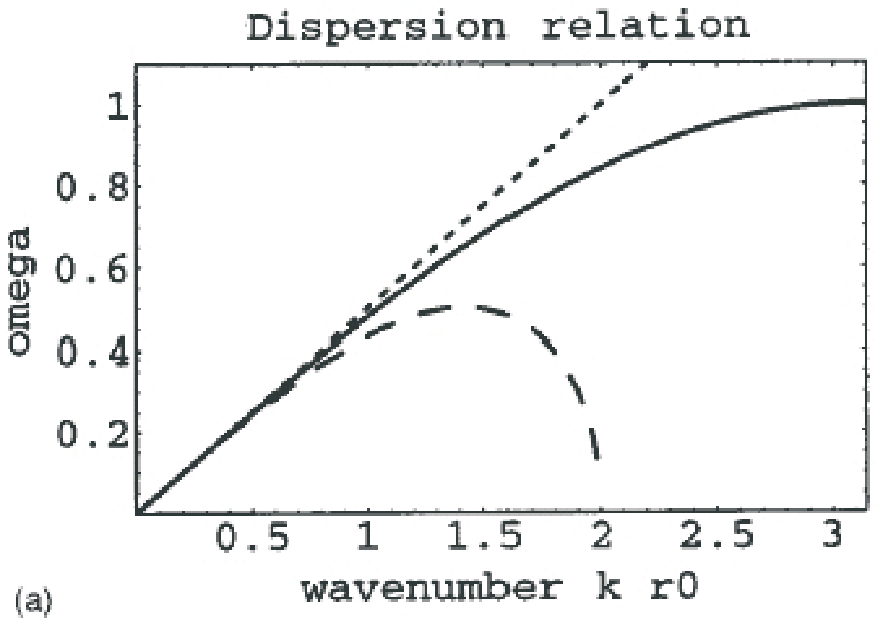}
} \caption{\small{The LDLW dispersion relation: frequency $\omega_L$ vs. wavenumber $k$ (solid curve). We have also depicted: the continuous approximation (dashed curve) and the acoustic (tangent) curve at the origin.}} \label{dispLDLW}
\end{figure}
The LDLW dispersion curve is depicted in Fig \ref{dispLDLW}.

The multiple scales (reductive
perturbation) technique (cf. above) now yields ($\sim \epsilon$)
%(to lowest order $\sim \epsilon$)
a {\em{zeroth-}}harmonic mode, describing a constant displacement,
viz. \[\delta x_n \approx \epsilon \, \bigl[u_0^{(1)} + \,
(u_1^{(1)} \, e^{i \phi_n} + {\rm{c.c.}})\bigr]
 + \epsilon^2 \,
(u_2^{(2)} \, e^{2 i \phi_n} + {\rm{c.c.}} ) \, + ... \, . \] The
1st-order  amplitudes obey the coupled equations [6]:
\begin{eqnarray}
i \frac{\partial u_1^{(1)}}{\partial T} \, + \, P_L \,
\frac{\partial^2 u_1^{(1)}}{\partial X^2} \, + \,Q_0 \,
|u_1^{(1)}|^2 u_1^{(1)}\, + \frac{p_0 k^2}{2 \omega_L} \,
u_1^{(1)} \frac{\partial u_0^{(1)}}{\partial X} = 0\, ,
\label{NLSE01}
\\
\frac{\partial^2 u_0^{(1)}}{\partial X^2} \, = \,- \frac{p_0
k^2}{v_{g, L}^2 - \omega_{L, 0}^2 r_0^2} \, \frac{\partial
}{\partial X} \, |u_1^{(1)}|^2 \, , \qquad \qquad \qquad
\label{NLSE02a}
\end{eqnarray}
where $v_{g, L} = \omega_L'(k)$; $\{ X, T \}$ are \emph{slow}
variables (as above). The description involves the definitions:
$p_0 = - r_0^{3} U'''(r_0)/M \equiv 2 a_{20} r_0^3$ and $q_0 =
U''''(r_0) r_0^4/(2 M) \equiv 3 a_{30} r_0^4$ (both positive
quantities of similar order of magnitude for Debye interactions;
see in [4, 7]).
 Eqs. (\ref{NLSE01}), (\ref{NLSE02a}) may be combined into a closed
 equation, which is identical to Eq. (\ref{NLSE}) (for $A = u_1^{(1)}$,
here).
 Now, here $P = P_L
= \omega_{L}''(k)/2 < 0$, while the form of $Q >0$ ($< 0$)
\cite{IKPKSLDLW} prescribes stability (instability) at low (high)
$k$.
%The existence of the zeroth mode now results in
Envelope excitations are now \emph{asymmetric}, i.e. rarefactive
bright or compressive  dark envelope structures (see Figs.).
% 2, 3).
\begin{figure}[htb]
\vskip -.3 cm
 \centering
 \resizebox{13.8cm}{!}{
 \includegraphics[]{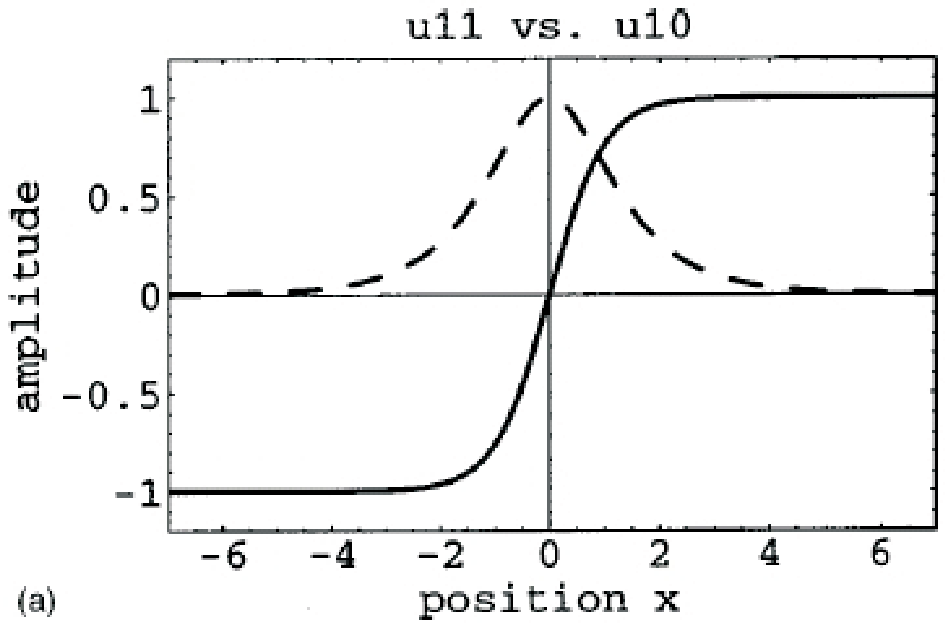}
\hskip 1 cm
\includegraphics{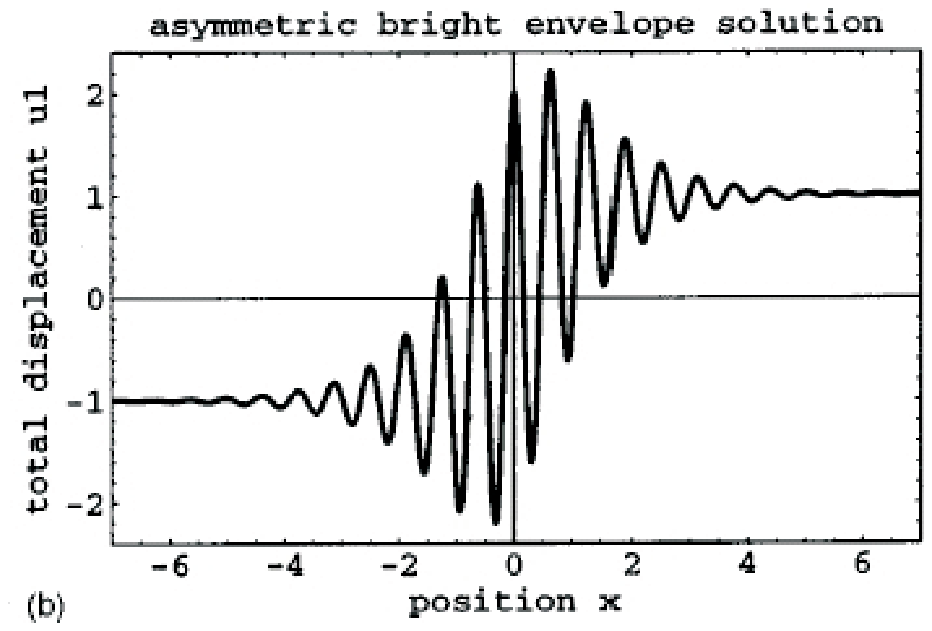}
}\vskip -.6 cm \caption{\small \emph{Bright} LDL
%packets
%modulated wavepackets
(asymmetric) envelope solitons: (a) the zeroth (pulse) and first
harmonic (kink) amplitudes; (b) the resulting asymmetric
wavepacket.} \label{fig2}
\end{figure}
\vskip -.7 cm
\begin{figure}[htb]
 \centering
 \resizebox{14.5cm}{!}{
 \includegraphics[]{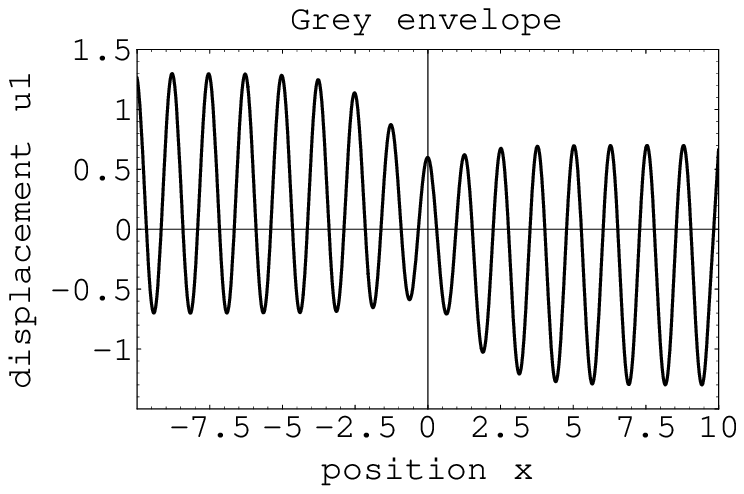}
\hskip 1 cm
\includegraphics{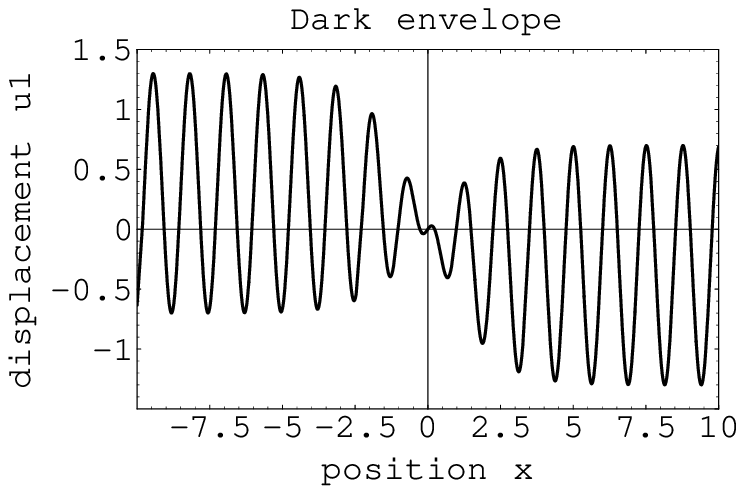}
}
%\vskip -.6 cm
\caption{\small (a) \emph{Grey} and (b) \emph{dark}
LDL (asymmetric) modulated wavepackets.} \label{fig3}
\end{figure}
%The kinetic coefficients involved in this description, $p_0 = -
%U'''(r_0) r_0^3/M$ and $q_0 = U''''(r_0) r_0^4/(2 M)$, are both
%positive for the Debye interactions.

\section{\emph{Longitudinal} solitons.} Equation
(\ref{discrete-eqmotion-x}) is identical to the equation of motion
in an atomic chain with anharmonic springs, i.e. in the celebrated
{\sc{FPU}} (\textit{Fermi-Pasta-Ulam}) problem. Inspired by
methods of solid state physics, one may opt for a continuum
description at a first step, viz. $\delta x_n(t) \rightarrow u(x,
t)$. This may lead to different nonlinear evolution equations
(depending on simplifying assumptions), some of which are
critically discussed in [9]. What follows is a summary of the
lengthy
%comparative
analysis carried out therein.

Keeping lowest order nonlinear and dispersive terms, the continuum
variable $u$ obeys \cite{comment1}:
\begin{equation}
\ddot{u}  \,+ \, \nu \, \dot{u} - c_L^2 \, u_{xx} -
\frac{c_{L}^2}{12}\, r_0^2 \, u_{xxxx} \, = \,- \, p_0
%2 \, a_{20}\, r_0^3
\, u_x \,u_{xx}
%\, + \, 2 \, a_{02}\, r_0^3 \, w_x \,w_{xx}
%\, -  \, a_{12}\, r_0^4 \, [(w_x)^2 \,u_{xx} + 2 w_x w_{xx} u_x ]
\, + \, q_0
%3 \, a_{30}\, r_0^4
\, (u_x)^2 \,u_{xx} \, , \label{eqmotion-gen-continuum-x}
\end{equation}
where $(\cdot)_x \equiv \partial (\cdot)/\partial x$; \ $c_L =
\omega_{L, 0} \, r_0$; $p_0$ and $q_0$ were defined above.
%the subscript denotes partial differentiation.
Assuming {\em{near-sonic propagation}} (i.e. $v \approx c_L$), and
defining the relative displacement $w = u_x$, one has
\begin{equation}
w_\tau \,- \, a\, w \, w_\zeta \,+ \, \hat a\, w^2 \, w_\zeta + \,
b\, w_{\zeta\zeta\zeta}\, = \, 0 \, \label{EKdV}
\end{equation}
(for $\nu = 0$), where $a= {p_0}/({2 c_L}) > 0$, $\hat a=
{q_0}/({2 c_L}) > 0$, and $b = {c_L r_0^2}/24 > 0$. Since the
original work of Melands\o \ [4], various studies have relied on
the
%(pre-existing knowledge on the)
{\em{Korteweg - deVries}} (KdV) equation, i.e. Eq. (\ref{EKdV})
for $\hat a=0$, in order to gain analytical insight in the
\emph{compressive} structures observed in experiments [1]. Indeed,
the KdV Eq. possesses {\em{negative}} ({\em{only}}, here, since $a
> 0$) supersonic pulse soliton solutions for $w$, implying a
compressive (anti-kink) excitation for $u$; the KdV soliton is
thus interpreted as a density variation in the crystal, viz. $n(x,
t)/n_0 \sim - \partial u/\partial x \equiv - w$. Also, the pulse
width $L_0$ and height $u_0$ satisfy $u_0 L_0^2 = cst.$, a feature
which is confirmed by experiments [1]. Now, here's a crucial point
to be made (among others [9]): in a Debye crystal, $\hat a \approx
2 a$ roughly (for $\kappa \approx 1$), so the KdV
%(lowest nonlinearity order)
approximation (i.e. assuming $\hat a \approx 0$) is not valid.
Instead, one may employ the \emph{extended KdV} Eq. (eKdV)
(\ref{EKdV}), which accounts for \emph{both} compressive
\emph{and} rarefactive lattice excitations (see expressions in
[9]; also cf. Fig. 4).
\begin{figure}[htb]
 \centering
 \resizebox{14.6cm}{!}{
 \includegraphics[]{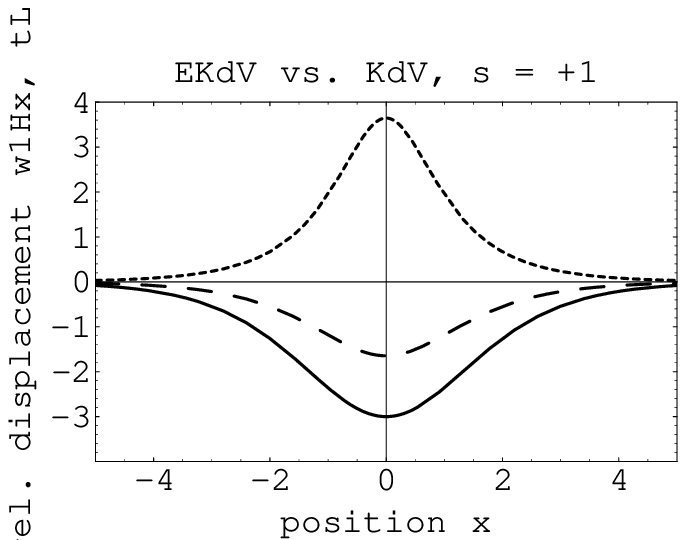}
\hskip 1 cm
\includegraphics{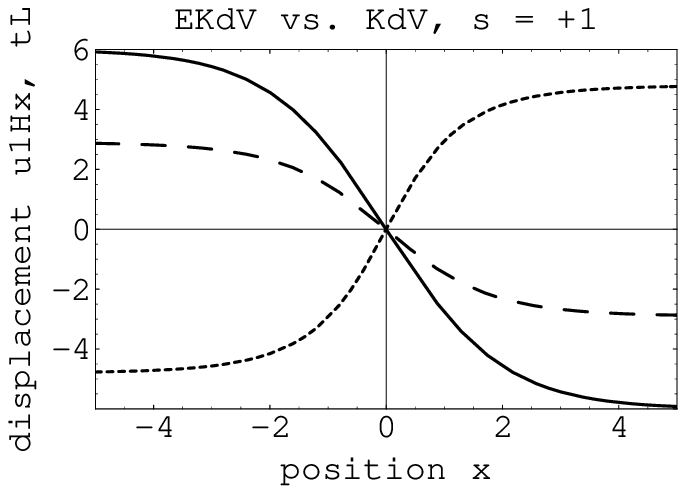}
} \caption{\small {Solutions of the \emph{extended} KdV Eq. (for
$q_0 > 0$; dashed curves) vs. those of the KdV Eq. (for $q_0 = 0$;
solid curves): (a) relative displacement $u_x$; (b) grain
displacement $u$.}} \label{fig4}
\end{figure}

Alternatively, Eq. (\ref{eqmotion-gen-continuum-x}) can be reduced
to
%the cast into
a \emph{Generalized Boussinesq} (GBq) Equation
\begin{equation}
\ddot{w}  \,- v_0^2 \, w_{xx}\,= h \, w_{xxxx}\, + \, p \,
(w^2)_{xx} \, + \, q \, (w^3)_{xx} \label{GB}
\end{equation}
($w = u_x$; $p = - p_0/2 < 0$, $q = q_0/3 > 0$); again, for $q
\sim q_0 = 0$, one recovers a \emph{Boussinesq} (Bq) equation,
e.g. widely studied in solid chains. As physically expected, the
GBq (Bq) equation yields, like its eKdV (KdV) counterpart, both
compressive and rarefactive (only compressive) solutions; however,
the (supersonic) propagation speed $v$ now does \emph{not} have to
be close to $c_L$. A detailed comparative study of (and exact
expressions for) all of these soliton excitations
%(related to existing soliton theories),
%which are directly derived from Eq. (\ref{eqmotion-gen-continuum-x}),
can be found in [9].
% and is too lengthy to reproduce here

\section{Conclusions.} \textit{Concluding}, we have reviewed
recent results on nonlinear excitations (solitary waves) occurring
in a (1d) dust mono-layer.
%One encounters e
Modulated envelope TDL and LDL structures occur, due to sheath and
coupling nonlinearity. Both compressive and rarefactive
longitudinal excitations are predicted and may be observed by
appropriate experiments.
% by the theory.

\renewcommand{\baselinestretch}{1.}

\small

\bigskip

\begin{acknowledgments}
This work was supported by the {\it{SFB591
(Sonderforschungsbereich) -- Universelles Verhalten
gleichgewichtsferner Plasmen: Heizung, Transport und
Strukturbildung}} German government Programme.
\end{acknowledgments}

%\noindent {\bf{Acknowledgements.}}

%\medskip \noindent {\bf{References}}


\begin{thebibliography}{10}
\bibitem{comment1} Only first neighbor interactions are considered
throughout this paper. See in [5] for details and coefficient
definitions.
\bibitem{comment2} Coupling anharmonicity, expressed by
a term $\sim \bigl[ (\delta z_{n+1} - \delta z_{n})^{3} - (\delta
z_{n} - \delta z_{n-1})^{3} \bigr] $, which is omitted in the
right-hand side of Eq. (\ref{eqmotion}), may be added at a later
stage.
\bibitem{comment3}
Follow exactly the definitions in [5, 6], not reproduced here.
\bibitem{comment4} The damping term is neglected by setting $\nu = 0$
in the following, thus omitting collisions with neutrals; for $\nu
\neq 0$, an imaginary part appears, in account for damping in both
dispersion relation $\omega(k)$ and the resulting envelope
equations.
\end{thebibliography}
\end{document}